\documentclass[12pt]{article}
\pdfoutput=1

\usepackage{amsmath}
\usepackage{amssymb}
\usepackage{epsfig}
\usepackage{epstopdf}
\usepackage{latexsym}
\usepackage{color}
\numberwithin{equation}{section}
\usepackage[
      colorlinks=false,
      linkcolor=darkblue,  
      urlcolor=blue,    
      filecolor=blue,     
      citecolor=red,
linktocpage=true,
      pdfstartview=FitV,
      bookmarksopen=true    
      ]{hyperref}

\usepackage{amssymb}
\usepackage[bbgreekl]{mathbbol}
\DeclareSymbolFontAlphabet{\mathbbl}{bbold}

\begin{document}

\begin{titlepage}

\centerline
\centerline
\centerline
\bigskip
\centerline{\Huge \rm Baryonic spindles from conifolds}
\bigskip 
\bigskip
\bigskip
\bigskip
\bigskip
\bigskip
\centerline{\rm Minwoo Suh}
\bigskip
\centerline{\it School of General Education, Kumoh National Institute of Technology,}
\centerline{\it Gumi, 39177, Korea}
\bigskip
\centerline{\tt minwoosuh1@gmail.com} 
\bigskip
\bigskip
\bigskip
\bigskip
\bigskip
\bigskip
\bigskip
\bigskip

\begin{abstract}
\noindent We construct supersymmetric $AdS_3\times\mathbbl{\Sigma}$ solutions with baryonic charge in the Betti-vector truncation of five-dimensional gauged $\mathcal{N}=4$ supergravity where $\mathbbl{\Sigma}$ is a spindle. The truncation is obtained from type IIB supergravity on $AdS_5\times{T}^{1,1}$. The solutions realize supersymmetry by the anti-twist. The dual field theories are 2d $\mathcal{N}=(0,2)$ SCFTs from the Klebanov-Witten theory compactified on a spindle. We calculate the holographic central charge of the solutions and it precisely matches the result from the gravitational block.
\end{abstract}

\vskip 3.5cm

\flushleft {April, 2023}

\end{titlepage}

\tableofcontents

\section{Introduction}

Topological twist, \cite{Witten:1988ze}, is a method to preserve supersymmetry which has benefited field theory and supergravity. In field theory, it was employed to obtain novel field theories in lower dimensions from their higher dimensional parent theories. In supergravity, novel $AdS$ solutions interpolating higher dimensional $AdS$ vacua were constructed, \cite{Maldacena:2000mw}.

Past several years, we have observed the discovery of two novel classes of supergravity solutions that preserve supersymmetry in ways different from the topological twist. The first class of solutions was obtained by wrapping branes on a spindle which is topologically a two-sphere with two conical deficit angles at two poles, \cite{Ferrero:2020laf}. The second class of the solutions is obtained by wrapping branes on a disk, topologically a two-sphere with a conical deficit angle, \cite{Bah:2021mzw, Bah:2021hei}.

These two classes of solutions have several novel features. First, as advertised, they preserve supersymmetry not by the topological twist. For the spindle solutions, there are two distinct classes of solutions by how they preserve supersymmetry: twist and anti-twist. If the spinors at the poles have an identical chirality, they are in the twist class. Otherwise, they are in the anti-twist class, \cite{Ferrero:2021etw, Couzens:2021cpk}. Second, unlike the Riemann surfaces employed for topological twists, the spindle and disk are orbifolds with non-constant curvature. Third, although the spindle and disk solutions share identical local solutions, they are physically distinct after their global completions. See \cite{Gutperle:2022pgw} for defect solutions from yet another global completion. Fourth, the singularities of the solutions in gauged supergravity are resolved when they are uplifted to supergravity in ten and eleven dimensions.

The spindle and disk solutions from various branes were constructed and studied. For the spindle solutions: from D3-branes, \cite{Ferrero:2020laf, Hosseini:2021fge, Boido:2021szx, Couzens:2022aki}, M2-branes, \cite{Ferrero:2020twa, Cassani:2021dwa, Ferrero:2021ovq, Couzens:2021rlk}, M5-branes, \cite{Ferrero:2021wvk}, D4-branes, \cite{Faedo:2021nub, Giri:2021xta}, mass-deformed D3-branes, \cite{Arav:2022lzo}, mass-deformed M2-branes, \cite{Suh:2022pkg}, and D2-branes, \cite{Couzens:2022yiv}. For the disk solutions: from M5-branes, \cite{Bah:2021mzw, Bah:2021hei, Karndumri:2022wpu, Couzens:2022yjl, Bah:2022yjf}, D3-branes, \cite{Couzens:2021tnv, Suh:2021ifj}, M2-branes, \cite{Suh:2021hef, Couzens:2021rlk}, and D4-branes, \cite{Suh:2021aik}. Then there are solutions from branes wrapped on four-dimensional orbifolds which are products of a spindle and another spindle or a Riemann surface: from M5-branes, \cite{Boido:2021szx, Suh:2022olh, Cheung:2022ilc}, and D4-branes, \cite{Giri:2021xta, Faedo:2021nub, Suh:2022olh, Couzens:2022lvg, Faedo:2022rqx}.

Although the field theory dual of M5-branes wrapped on a disk was proposed, \cite{Bah:2021mzw, Bah:2021hei, Couzens:2022yjl, Bah:2022yjf}, to be a class of Argyres-Douglas theories, \cite{Argyres:1995jj}, other solutions obtained from disks and spindles have not been identified.

In this paper, we construct spindle solutions with baryonic charges. In particular, we consider five-dimensional gauged $\mathcal{N}=4$ supergravity from type IIB supergravity, \cite{Schwarz:1983qr, Howe:1983sra}, on $AdS_5\times{T}^{1,1}$, \cite{Cassani:2010na, Bena:2010pr, Halmagyi:2011yd}. Via the AdS/CFT correspondence, \cite{Maldacena:1997re}, they are dual to the Klebanov-Witten theories, an infinite family of quiver gauge theories by the worldvolume gauge theory living on a stack of N D3 branes probing the tip of a three-dimensional Calabi-Yau cone with a five-dimensional Sasaki Einstein base, $T^{1,1}$ in this case, \cite{Klebanov:1998hh}. Unlike five-sphere, $T^{1,1}$ is diffeomorphic to $S^2\times{S}^3$ and has non-vanishing second Betti-number. The non-trivial two-cycle introduces a Betti vector in the truncation to five-dimensional gauged supergravity. The Betti vector is dual to the baryonic flavor symmetry in field theory. 

Recently, a systematic method to calculate the supersymmetric action of $AdS$ solutions, including the spindle solutions, by extremizing the gravitational blocks, \cite{Hosseini:2019iad}, has been developed, \cite{Boido:2022mbe}. See \cite{Hosseini:2021fge, Faedo:2021nub, Faedo:2022rqx, Boido:2022iye} for the earlier works on the gravitational blocks for spindle solutions. The supersymmetric action agrees with the $c$-functions in the dual field theory. Although this method enables us to calculate the $c$-function without the knowledge of explicit solutions in supergravity, the $c$-function from this calculation does not guarantee the existence of the corresponding solution. Thus, our solution explicitly tests the gravitational block calculation for the spindle solutions with baryonic charges presented in \cite{Boido:2022mbe}.

To construct the spindle solutions with Betti vector multiplet and also with hypermultiplets, we employ the method developed in \cite{Arav:2022lzo} for spindle solutions with hypermultiplets. Furthermore, for the structure of the work, we will closely follow \cite{Arav:2022lzo} as it is well organized and also facilitates the comparison.

In section \ref{sec2}, we review a subtruncation of the Betti-vector truncation. In section \ref{sec3}, we study the BPS equations and calculate the holographic central charge. In section \ref{sec4}, we numerically construct the solutions. In section \ref{sec5}, the gravitational block calculation for the central charge is presented, and the result precisely matches the supergravity calculation. In section \ref{sec6}, we conclude. In appendix \ref{appA}, we review the details of Betti-vector truncation. In appendix \ref{appB}, we present the derivation of the BPS equations.

\bigskip

\noindent {\bf Note added}: After submitting the manuscript to arXiv, we learned that A. Amariti, N. Petri, and A. Segati have worked on the same problem, \cite{Amariti:2023mpg}.

\section{The supergravity model} \label{sec2}

We consider a subtruncation of the Betti-vector truncation in \cite{Halmagyi:2011yd, Amariti:2016mnz} which is reviewed in appendix \ref{appA}. The field content consists of the graviton, three vector fields, $A^0$, $A^1$, $A^2$, two real scalar fields, $u_2$, $u_3$, from vector multiplets, and a complex scalar field, $(u_1,\theta)$, from hypermultiplets which are charged by the combination of vector fields, $2A^0-A^1-A^2$.

The bosonic Lagrangian of the truncation in a mostly plus signature is given by
\begin{align} \label{bvlag}
e^{-1}\mathcal{L}\,=\,&R-8\partial_\mu{u}_1\partial^\mu{u}_1-\frac{1}{2}e^{-8u_1}D_\mu\theta{D}^\mu\theta-4\partial_\mu{u}_2\partial^\mu{u}_2-12\partial_\mu{u}_3\partial^\mu{u}_3-V \notag \\
&-\frac{1}{4}e^{-8u_3}F^0_{\mu\nu}F^{0\mu\nu}-\frac{1}{4}e^{-4u_2+4u_3}F^1_{\mu\nu}F^{1\mu\nu}-\frac{1}{4}e^{4u_2+4u_3}F^2_{\mu\nu}F^{2\mu\nu}+\frac{1}{4}\epsilon^{\mu\nu\rho\sigma\delta}F^1_{\mu\nu}F^2_{\rho\sigma}A^0_\delta\,,
\end{align}
and we define
\begin{equation}
D\theta\,\equiv\,d\theta-4A^0+2A^1+2A^2\,.
\end{equation}
The scalar potential is given by
\begin{equation}
V\,=\,\frac{1}{8}\left(\frac{\partial{W}}{\partial{u_1}}\right)^2+\frac{1}{4}\left(\frac{\partial{W}}{\partial{u_2}}\right)^2+\frac{1}{12}\left(\frac{\partial{W}}{\partial{u_3}}\right)^2-\frac{4}{3}W^2\,,
\end{equation}
where the superpotential is
\begin{equation} \label{superpotentialw}
W\,=\,\left(e^{2u_2-2u_3}+e^{-2u_2-2u_3}-2e^{4u_3}\right)e^{-4u_1}+3e^{4u_3}\,.
\end{equation}
The scalar potential admits an $\mathcal{N}=2$ supersymmetric $AdS_5$ vacuum with a unit radius located where the scalar fields are trivial. The Betti vector, $A_{\mathcal{B}}$, the massless vector appearing in the Sasaki-Einstein truncation, $A_g$, and the massive vector, $A_m$, are given by
\begin{equation}
A_{\mathcal{B}}\,=\,\frac{A^2-A^1}{\sqrt{2}}\,, \qquad A_m\,=\,\frac{A^1+A^2-2A^0}{\sqrt{6}}\,, \qquad A_g\,=\,\frac{A^0+A^1+A^2}{\sqrt{3}}\,,
\end{equation}
and, when $A^1\,=\,A^2$, the Betti vector vanishes and after a suitable truncation of the scalar fields, it returns to the Sasaki-Einstein truncation. There is a $\mathbb{Z}_2$ symmetry of the model,
\begin{equation} \label{u2a1a2}
u_2\,\rightarrow\,-u_2\,, \qquad A^1\,\leftrightarrow\,A^2\,.
\end{equation}

The supersymmetry variations of gravitino, gaugino, and hyperino reduce to
\begin{align}
\left[6\nabla_\mu-6iQ_\mu-W\gamma_\mu-\frac{i}{2}H_{\nu\rho}\left(\gamma_\mu\,^{\nu\rho}-4\delta_\mu^\nu\gamma^\rho\right)\right]\epsilon\,=\,0\,, \notag \\
\Big[8\partial_\mu{u}_1\gamma^\mu+\partial_{u_1}W+2i\partial_{u_1}Q_\mu\gamma^\mu\Big]\epsilon\,=\,0\,, \notag \\
\left[4\partial_\mu{u}_2\gamma^\mu+\partial_{u_2}W+\frac{i}{2}\partial_{u_2}H_{\mu\nu}\gamma^{\mu\nu}\right]\epsilon\,=\,0\,, \notag \\
\left[12\partial_\mu{u}_3\gamma^\mu+\partial_{u_3}W+\frac{i}{2}\partial_{u_3}H_{\mu\nu}\gamma^{\mu\nu}\right]\epsilon\,=\,0\,,
\end{align}
and we have
\begin{align} \label{hqdef}
H_{\mu\nu}\,=&\,-\frac{1}{2}\left(e^{-4u_3}F^0_{\mu\nu}+e^{-2u_2+2u_3}F^1_{\mu\nu}+e^{2u_2+2u_3}F^2_{\mu\nu}\right)\,, \notag \\
Q_\mu\,=&\,-\frac{1}{2}\Big[\left(3-2e^{-4u_1}\right)A^0_\mu+e^{-4u_1}\left(A^1_\mu+A^2_\mu\right)\Big]+\frac{1}{2}e^{-4u_1}D_\mu\theta\,.
\end{align}
We emphasize that the supersymmetry parameter is charged by $Q_\mu$ and the gauge fields in $Q_\mu$ are dressed by the scalar field, $u_1$. This is a distinct feature from the truncations with no baryonic charges studied for spindle solutions in the literature. The R-symmetry gauge field is given by
\begin{equation}
A^{\text{R}}_\mu\,\equiv\,A^0_\mu+A^1_\mu+A^2_\mu\,,
\end{equation}
and when the complex scalar field, $(u_1,\theta)$, is vanishing, we find $A^{\text{R}}_\mu=-2Q_\mu$.

\section{$AdS_3$ ansatz} \label{sec3}

We consider the ansatz for the metric and the gauge fields, \cite{Arav:2022lzo},
\begin{align}
ds^2\,=&\,e^{2V}ds^2_{AdS_3}+f^2dy^2+h^2dz^2\,, \notag \\
A^I\,=&\,a^Idz\,,
\end{align}
where $V$, $f$, $h$, and $a^I$, $I=0,1,2$, are functions of the coordinate $y$ only. In order to avoid partial differential equations from the equations of motion for gauge fields, the scalar field, $\theta$, is given by $\theta\,=\,\bar{\theta}z$ where $\bar{\theta}$ is a constant. Then we find
\begin{equation}
Q_\mu{d}x^\mu\,\equiv\,Q_zdz\,,
\end{equation}
where $Q_z$ is a function of the coordinate $y$. All other scalar fields are functions of the coordinate $y$ only.

We choose the orthonormal frame,
\begin{equation} \label{orthobasis}
e^a\,=\,e^V\bar{e}^a\,, \qquad e^3\,=\,fdy\,, \qquad e^4\,=\,hdz\,,
\end{equation}
where $\bar{e}^a$ is an orthonormal frame on $ds^2_{AdS_3}$. The frame components of the field strengths are 
\begin{equation}
F^I_{34}\,=\,f^{-1}h^{-1}\left(a^I\right)'\,.
\end{equation}

From the equations of motion for gauge fields, we find the integrals of motion,
\begin{align} \label{eereef}
-\frac{1}{2}e^{3V}\left(e^{-4u_2+4u_3}F^1_{34}-e^{4u_2+4u_3}F^2_{34}\right)\,=&\,\mathcal{E}_\mathcal{B}\,, \notag \\
\frac{1}{2}e^{3V}\left(e^{-8u_3}F^0_{34}+e^{-4u_2+4u_3}F^1_{34}+e^{4u_2+4u_3}F^2_{34}\right)\,=&\,\mathcal{E}_R\,, \notag \\
\left(e^{3V-8u_3}F^0\right)'\,=&\,e^{3V}fh^{-1}4e^{-8u_1}D_z\theta\,,
\end{align}
where $\mathcal{E}_\mathcal{B}$ and $\mathcal{E}_R$ are constant and $D_z\theta=\left(\bar{\theta}-4a^0+2a^1+2a^2\right)$.

\subsection{BPS equations}

We employ the gamma matrices of mostly plus signature,
\begin{equation} \label{ansatzads3}
\gamma^m\,=\,\Gamma^m\otimes\sigma^3\,, \qquad \gamma^3\,=\,1\otimes\sigma^1\,, \qquad \gamma^4\,=\,1\otimes\sigma^2\,,
\end{equation}
where $\Gamma^m=\left(i\sigma^2,\sigma^3,\sigma^1\right)$ are three-dimensional gamma matrices. The Killing spinors are given by
\begin{equation}
\epsilon\,=\,\psi\otimes\chi\,,
\end{equation}
and the three-dimensional spinor on $AdS_3$ satisfies
\begin{equation}
D_m\psi\,=\,\frac{1}{2}\kappa\Gamma_m\psi\,,
\end{equation}
where $\kappa\,=\,\pm1$ fixes the chirality. The spinor, $\chi$, is given by 
\begin{equation}
\chi\,=\,e^{V/2}e^{isz}\left(
\begin{array}{ll}
 \sin\frac{\xi}{2} \\
 \cos\frac{\xi}{2}
\end{array}
\right)\,,
\end{equation}
where the constant, $s$, is the gauge dependent charge under the action of azimuthal Killing vector, $\partial_z$.

From the derivation given in appendix \ref{appB}, for $\sin\xi\ne0$, the BPS equations are
\begin{align} \label{bpsequations}
f^{-1}\xi'\,=&\,W\cos\xi+2\kappa{e}^{-V}\,, \notag \\
f^{-1}V'\,=&\,\frac{1}{3}W\sin\xi\,, \notag \\
f^{-1}u_1'\,=&\,-\frac{1}{8}\frac{\partial_{u_1}W}{\sin\xi}\,, \notag \\
f^{-1}u_2'\,=&\,-\frac{1}{4}\partial_{u_2}W\sin\xi\,, \notag \\
f^{-1}u_3'\,=&\,-\frac{1}{12}\partial_{u_3}W\sin\xi\,, \notag \\
f^{-1}\frac{h'}{h}\,=&\,\frac{1}{\sin\xi}\left(2\kappa{e}^{-V}\cos\xi+\frac{W}{3}\left(1+2\cos^2\xi\right)\right)\,,
\end{align}
with two constraints,
\begin{align} \label{twotwocon}
\left(s-Q_z\right)\sin\xi\,=&\,-\frac{1}{2}Wh\sin\xi-\kappa{h}e^{-V}\,, \notag \\
\frac{1}{2}\partial_{u_1}\cos\xi\,=&\,\partial_{u_1}Q_z\sin\xi{h}^{-1}\,.
\end{align}
The field strengths of the gauge fields are given by
\begin{align} \label{wfrel}
e^{-4u_3}F^0_{34}\,=&\,\frac{1}{3}\left(2W+\partial_{u_3}W\right)\cos\xi+2\kappa{e}^{-V}\,, \notag \\
e^{-2u_2+2u_3}F^1_{34}\,=&\,\frac{1}{6}\left(4W+3\partial_{u_2}W-\partial_{u_3}W\right)\cos\xi+2\kappa{e}^{-V}\,, \notag \\
e^{2u_2+2u_3}F^2_{34}\,=&\,\frac{1}{6}\left(4W-3\partial_{u_2}W-\partial_{u_3}W\right)\cos\xi+2\kappa{e}^{-V}\,.
\end{align}
We have checked that the BPS equations solve the equations of motion from the Lagrangian, \eqref{bvlag}.

\subsection{Integrals of motion}

There is an integral of the BPS equations,
\begin{equation}
he^{-V}\,=\,k\sin\xi\,,
\end{equation}
where $k$ is a constant. The poles of the metric is where we have $h=0$. Hence, at the poles, we also have $\sin\xi=0$. From \eqref{bpsequations} and \eqref{twotwocon}, we obtain
\begin{equation}
\xi'\,=\,-2k^{-1}\left(s-Q_z\right)\left(e^{-V}f\right)\,,
\end{equation}
and the two constraints in \eqref{twotwocon} are written by
\begin{align} \label{constraintstwo}
\left(s-Q_z\right)\,=&\,-k\left[\frac{1}{2}We^V\cos\xi+\kappa\right]\,, \notag \\
\frac{1}{2}\partial_{u_1}W\cos\xi\,=&\,k^{-1}e^{-V}\partial_{u_1}Q_z\,.
\end{align}

Using \eqref{wfrel}, the integrals of motion in \eqref{eereef} are given by
\begin{align} \label{eref}
\mathcal{E}_R\,=&\,e^{2V}\left[3e^V\cos\xi+\kappa\left(e^{-4u_3}+2e^{2u_3}\cosh\left(2u_2\right)\right)\right]\,, \notag \\
\mathcal{E}_\mathcal{B}\,=&\,2\kappa{e}^{2V}e^{2u_3}\sinh\left(2u_2\right)\,.
\end{align}

\subsection{Boundary conditions for spindle solutions} \label{sec33}

Following the analysis in \cite{Arav:2022lzo}, we study the boundary conditions for spindle solutions in this section. We fix the metric to be in conformal gauge,
\begin{equation} \label{confgauge}
f\,=\,-e^V\,,
\end{equation}
and the metric is given by
\begin{equation}
ds^2\,=\,e^{2V}\left[ds_{AdS_3}^2+ds_{\mathbbl{\Sigma}}^2\right]\,,
\end{equation}
where we have
\begin{equation}
ds_{\mathbbl{\Sigma}}^2\,=\,dy^2+k^2\sin^2\xi{d}z^2\,,
\end{equation}
for the metric on a spindle, ${\mathbbl{\Sigma}}$. For the spindle solutions, there are two poles at $y=y_{N,S}$ with deficit angles of $2\pi\left(1-\frac{1}{n_{N,S}}\right)$. The azimuthal angle, $z$, has a period which we set
\begin{equation}
\Delta{z}\,=\,2\pi\,.
\end{equation}

\subsubsection{Analysis of the BPS equations}

At the poles, we have $\sin\xi\rightarrow0$ and thus we find $\cos\xi\rightarrow\pm1$. We introduce $t_{N,S}\in\{0,1\}$ for $\cos\xi_{N,S}=(-1)^{t_{N,S}}$. The poles are at $y=y_{N,S}$ and we choose them to be $y_N<y_S$ and $y\in[y_N,y_S]$. The deficit angles at the poles are $2\pi\left(1-\frac{1}{n_{N,S}}\right)$ with $n_{N,S}>1$. Thus, for the metric, we should have $|\left(k\sin\xi\right)'|_{N,S}=\frac{1}{n_{N,S}}$. From the symmetry of the BPS equations, \eqref{hsymm}, we choose{\footnote{We have different choices of signs for $f$ and $h$ in \eqref{confgauge} and \eqref{hzero} from \cite{Arav:2022lzo}.}}
\begin{equation} \label{hzero}
h\le0\,, \qquad \Leftrightarrow \qquad k\sin\xi\le0\,.
\end{equation}
Then we have $(k\sin\xi)'|_N<0$ and $(k\sin\xi)'|_S>0$. Thus, for $y\in[y_N,y_S]$, we impose
\begin{equation}
\left(k\sin\xi\right)'|_{N,S}\,=\,-\frac{(-1)^{l_{N,S}}}{n_{N,S}}\,, \qquad l_N=0\,, l_S\,=\,1\,.
\end{equation}

There are two different classes of solutions,
\begin{align} \label{twistantitwist}
\cos\xi|_{N,S}\,=\,(-1)^{t_{N,S}}; \qquad &\text{Twist:} \,\, \qquad \quad \left(t_N,t_S\right)\,=\,\left(1,1\right) \quad \text{or} \quad \left(0,0\right)\,, \notag \\
&\text{Anti-Twist:} \quad \left(t_N,t_S\right)\,=\,\left(1,0\right) \quad \text{or} \quad \left(0,1\right)\,,
\end{align}
where the spinors have identical or opposite chiralities at the poles for the twist and anti-twist classes, respectively.

As we have $(k\sin\xi)'=+2\cos\xi(s-Q_z)$ from the BPS equations in \eqref{bpsequations} and \eqref{confgauge}, we obtain
\begin{equation}
\left(s-Q_z\right)|_{N,S}\,=\,\frac{1}{2n_{N,S}}(-1)^{l_{N,S}+t_{N,S}+1}\,.
\end{equation}
From the BPS equation for $u_1$ in \eqref{bpsequations}, at the poles where we have $\sin\xi=0$, $\partial_{u_1}W$ should vanish to have a finite value of $u_1$. Then,  as $\partial_{u_1}W|_{N,S}=0$, from the second constraint in \eqref{constraintstwo}, we require $\partial_{u_1}Q_z|_{N,S}=0$, 
\begin{equation} \label{dqdwzero}
\partial_{u_1}Q_z|_{N,S}\,=\,\partial_{u_1}W|_{N,S}\,=\,0\,.
\end{equation}
Thus we find a constraint,{\footnote{As it was emphasized around \eqref{hqdef}, the combination of the gauge fields in $Q_\mu$ is dressed with the scalar field, $u_1$: this is a distinct feature from the truncations with no baryonic charges studied in the literature for spindle solutions and we cannot simply follow \cite{Arav:2022lzo} for the analysis in this paragraph. The spindle twist condition will be derived later in \eqref{spindletwist}.

Even for the topologically twisted solutions of $AdS_3\times\Sigma_{\mathfrak{g}}$ where $\Sigma_{\mathfrak{g}}$ is a Riemann surface with genus $\mathfrak{g}$, the twist condition depends on the scalar field, $u_1$, \cite{Amariti:2016mnz},
\begin{equation}
\left(3-2e^{-4u_1}\right)a_0+e^{-4u_1}\left(a_1+a_2\right)\,=\,-\kappa\,,
\end{equation}
where $a^I$, $I=0,1,2$, are constant magnetic charges and $\kappa=\pm1$ is the curvature of the Riemann surface. Thus it reduces to two constraints,
\begin{equation}
2a^0-a^1-a^2\,=\,0\,, \qquad a^0\,=\,-\frac{\kappa}{3}\,.
\end{equation}
See \cite{Bobev:2014jva} also for the case of Leigh-Strassler compactified on a Riemann surface. We would like to thank Jerome Gauntlett for the helpful discussion on this.}}
\begin{equation}
0\,=\,\partial_{u_1}Q_z|_{N,S}\,=\,\left[-2e^{-4u_1}\left(2A^0-A^1-A^2\right)-2e^{-4u_1}D_z\theta\right]|_{N,S}\,,
\end{equation}
which implies{\footnote{This condition appears to be a generalization of $\left[D_z\theta\right]|_{N,S}=0$ for the case with no baryonic charge, \cite{Arav:2022lzo}.}}
\begin{equation}
\left[D_z\theta+\left(2A^0-A^1-A^2\right)\right]|_{N,S}\,=\,0\,.
\end{equation}
Then, from \eqref{dtheta}, we deduce that
\begin{equation} \label{bartheta}
\bar{\theta}\,=\,\left(2A^0-A^1-A^2\right)|_{N,S}\,.
\end{equation}
We find the flux of the $U(1)$ gauge field by which the complex scalar field is charged is vanishing,
\begin{equation}
\frac{1}{2\pi}\int_{\mathbbl{\Sigma}}\left(2F^0-F^1-F^2\right)\,=\,\left(2A^0-A^1-A^2\right)|_N^S\,=\,\bar{\theta}-\bar{\theta}\,=\,0\,,
\end{equation}
where we used \eqref{bartheta}.

From the second condition in \eqref{dqdwzero} and the expression of the superpotential in \eqref{superpotentialw}, we find
\begin{equation} \label{dwzero}
\left(e^{6u_3}-\cosh\left(2u_2\right)\right)|_{N,S}\,=\,0\,, \qquad \Rightarrow \qquad W|_{N,S}\,=\,3e^{4u_3}|_{N,S}\,.
\end{equation}
We introduce two quantities,
\begin{equation} \label{defm1m2}
M_{(1)}\,\equiv\,-e^{4u_3}e^V\,, \qquad M_{(2)}\,\equiv\,-\frac{1}{2}\kappa-\frac{3}{2}M_{(1)}\cos\xi\,,
\end{equation}
and note that $M_{(1)}<0$. Then the integrals of motion in \eqref{eref} can be written by
\begin{align} \label{ereferef}
-\frac{1}{2}\mathcal{E}_R\,=&\,M_{(1)}^2\left[-\kappa+M_{(2)}e^{-12u_3}\right]+\kappa{e}^{2V+2u_3}\left(e^{6u_3}-\cosh\left(2u_2\right)\right)\,, \notag \\
\frac{1}{4}\left(\mathcal{E}_\mathcal{B}\right)^2\,=&\,M_{(1)}^4\left[1-e^{-12u_3}\right]+M_{(1)}^4e^{-12u_3}\left[\cosh^2\left(2u_2\right)-e^{12u_3}\right]\,,
\end{align}
and note that the second terms on the right-hand sides vanish at the poles for \eqref{dwzero}. Also from \eqref{dwzero} and the first constraint in \eqref{constraintstwo}, we obtain that, at the poles,
\begin{align} \label{m1m2m1m2}
M_{(1)}|_{N,S}\,=&\,\frac{1}{3}\left[2(-1)^{t_{N,S}}\kappa-\frac{1}{kn_{N,S}}(-1)^{l_{N,S}}\right]\,, \notag \\
M_{(2)}|_{N,S}\,=&\,\frac{1}{4}\left[2\kappa-\frac{2}{kn_{N,S}}(-1)^{l_{N,S}+t_{N,S}}\right]\,.
\end{align}
Then we can determine the value of the scalar field, $u_3$, at the poles in terms of the spindle numbers, $(n_{N,S},t_{N,S},k)$, by solving the set of linear equations,
\begin{equation} \label{matrixeq}
\left(
\begin{array}{ll}
\,\,\,\,\,\, -M_{(1)}^4|_N & \,\,\,\,\,\,\,\,\, M_{(1)}^4|_S \\
M_{(1)}^2|_NM_{(2)}|_N & -M_{(1)}^2|_SM_{(2)}|_S
\end{array}
\right)\left(
\begin{array}{ll}
e^{-12u_{3N}} \\
e^{-12u_{3S}}
\end{array}
\right)\,=\,\left(
\begin{array}{ll}
\,\,\,\,\,\, M_{(1)}^4|_S-M_{(1)}^4|_N \\
-\kappa{M}_{(1)}^4|_S+\kappa{M}_{(1)}^4|_N
\end{array}
\right)\,.
\end{equation}
The values of the scalar field, $u_3$, at the poles are 
\begin{align} \label{poleu3}
e^{-12u_{3N}}\,=\,\frac{\left(M_{(1)}^2|_N-M_{(1)}^2|_S\right)\left[\left(M_{(1)}^2|_N+M_{(1)}^2|_S\right)M_{(2)}|_S-\kappa{M}_{(1)}^2|_S\right]}{M_{(1)}^2|_N\left(M_{(1)}^2|_NM_{(2)}|_S-M_{(1)}^2|_SM_{(2)}|_N\right)}\,, \notag \\
e^{-12u_{3S}}\,=\,\frac{\left(M_{(1)}^2|_N-M_{(1)}^2|_S\right)\left[\left(M_{(1)}^2|_N+M_{(1)}^2|_S\right)M_{(2)}|_N-\kappa{M}_{(1)}^2|_N\right]}{M_{(1)}^2|_S\left(M_{(1)}^2|_NM_{(2)}|_S-M_{(1)}^2|_SM_{(2)}|_N\right)}\,.
\end{align}
Furthermore, we determine the values of $u_2$ and $V$ at the poles from \eqref{dwzero} and the definition of $M_{(1)}$ in \eqref{defm1m2},
\begin{align} \label{poleu2V}
u_2|_{N,S}\,=&\,\pm\frac{1}{2}\text{Cosh}^{-1}\left(e^{6u_3}\right)|_{N,S}\,, \notag \\
V|_{N,S}\,=&\,\log\left(-e^{-4u_3}M_{(1)}\right)|_{N,S}\,.
\end{align}
Also from \eqref{ereferef} we must have
\begin{equation} \label{u301}
0\,<\,e^{-12u_{3N,S}}\,\le1\,,
\end{equation}
which restricts the range of $k$ for given spindle numbers, $n_{N,S}$ and $t_{N,S}$. From \eqref{m1m2m1m2}, the values of $u_2$, $u_3$ and $V$ at the poles depend only on $kn_{N,S}$.{\footnote {From $\mathcal{E}_\mathcal{B}$ in \eqref{eref} the sign of $\mathcal{E}_\mathcal{B}$ is the sign of $u_2\kappa$. Thus for any solution of $u_{3N,S}$ there are two boundary conditions of $u_2$ for a choice of $\kappa$.}} 

In this section, we have determined the values of the scalar fields, $u_2$ and $u_3$, and the metric function, $V$, in \eqref{poleu3} and \eqref{poleu2V}, in terms of the spindle data, $n_N$, $n_S$, $t_{N,S}$. In the end, we only have to specify the value of the scalar field, $u_1$, at the poles.

\subsubsection{Fluxes}

In \eqref{introi1i2i3} in appendix \ref{appB}, we find the expression of the field strengths,
\begin{equation}
F^I_{yz}\,=\,\left(a^I\right)'\,=\,\left(\mathcal{I}^I\right)'\,,
\end{equation}
with
\begin{equation}
\mathcal{I}^{(0)}\,\equiv\,-ke^V\cos\xi{e}^{4u_3}\,, \qquad \mathcal{I}^{(1)}\,\equiv\,-ke^V\cos\xi{e}^{2u_2-2u_3}\,, \qquad \mathcal{I}^{(2)}\,\equiv\,-ke^V\cos\xi{e}^{-2u_2-2u_3}\,.
\end{equation}
Then the fluxes are expressed by the pole data,
\begin{equation}
\frac{p_I}{n_Nn_S}\,\equiv\,\frac{1}{2\pi}\int_{\mathbbl{\Sigma}}F^I\,=\,\mathcal{I}^{(I)}|_N^S\,.
\end{equation}

Furthermore, the expressions can be written in terms of $M_{(1)}$,
\begin{align}
\mathcal{I}^{(0)}|_{N,S}\,=&\,\mathcal{I}_0|_{N,S}\,, \notag \\
\mathcal{I}^{(1)}|_{N,S}\,=&\,\left(\mathcal{I}_0\pm\mathcal{I}_\Delta\right)|_{N,S}\,, \notag \\
\mathcal{I}^{(2)}|_{N,S}\,=&\,\left(\mathcal{I}_0\mp\mathcal{I}_\Delta\right)|_{N,S}\,,
\end{align}
where we have
\begin{align} \label{i0id}
\mathcal{I}_0|_{N,S}\,\equiv&\,kM_{(1)}|_{N,S}(-1)^{t_{N,S}}\,, \notag \\
\mathcal{I}_\Delta|_{N,S}\,\equiv&\,kM_{(1)}|_{N,S}(-1)^{t_{N,S}}\sqrt{1-e^{-12u_{3N,S}}}\,,
\end{align}
and the $\pm$ is related to the sign choice for $u_2$. Finally, we show that the R-symmetry flux gives the twist and anti-twist conditions for the spindle solutions and massive vector flux vanishes, respectively,
\begin{align} \label{spindletwist}
\mathcal{I}_0|_N^S\,=&\,\frac{1}{3}\left(\mathcal{I}^{(0)}+\mathcal{I}^{(1)}+\mathcal{I}^{(2)}\right)|_N^S\,=\,\frac{1}{3}\frac{n_N(-1)^{t_S+1}+n_S(-1)^{t_N+1}}{n_Nn_S}\,, \notag \\
&\left(2\mathcal{I}^{(0)}-\mathcal{I}^{(1)}-\mathcal{I}^{(2)}\right)|_{N,S}\,=\,0\,.
\end{align}
The flux quantization conditions of three gauge fields imply that
\begin{enumerate}
\item $p_0=n_Nn_S\mathcal{I}_0|_N^S\in\mathbb{Z}$. This condition depends only on $n_N$, $n_S$ and is true while $n_N(-1)^{t_S}+n_S(-1)^{t_N}$ is multiple of 3.
\item We define  $p_\mathcal{B}\equiv\frac{{p}_1-p_2}{2}=\text{sign}(u_2)n_Nn_S\mathcal{I}_\Delta|_N^S\in\mathbb{Z}$. This gives a condition on $k$.
\item $2p_0=2p_1-2p_\mathcal{B}$ and thus $p_\mathcal{B}$ is an integer.
\end{enumerate}
From \eqref{i0id} we find
\begin{equation}
\mathcal{I}_\Delta|_{N,S}\,=\,\frac{1}{2}k\frac{(-1)^{t_{N,S}}}{M_{(1)}|_{N,S}}|\mathcal{E}_\mathcal{B}|\,,
\end{equation}
where $\mathcal{E}_\mathcal{B}$ is given in \eqref{eref}. As $\mathcal{E}_\mathcal{B}$ is conserved and has identical values at the poles, we obtain
\begin{equation}
\mathcal{I}_\Delta|_N^S\,=\,\frac{1}{2}k|\mathcal{E}_\mathcal{B}|\left[\frac{(-1)^{t_S}}{M_{(1)}|_S}-\frac{(-1)^{t_N}}{M_{(1)}|_N}\right]\,=\,\frac{1}{2}|\mathcal{E}_\mathcal{B}|\frac{(-1)^{t_N+t_S+1}}{M_{(1)}|_SM_{(1)}|_N}\left(\mathcal{I}_0|_N^S\right)\,,
\end{equation}
and hence
\begin{equation}
\frac{p_\mathcal{B}^2}{p_0^2}\,=\,\frac{\left(M_{(1)}|_N\right)^2}{\left(M_{(1)}|_S\right)^2}\left(1-e^{-12u_{3N}}\right)\,.
\end{equation}
Note that $\text{sign}(u_2)=\text{sign}(p_{\mathcal{B}})\text{sign}(p_0)(-1)^{t_N+t_S+1}$. From the values of $M_{(1)}$ and $e^{-12u_3}$ at the poles in terms of $n_N$, $n_S$, $t_S$ and $k$ by solving \eqref{matrixeq}, we invert the expression to solve for $k$. For the twist class, labeled by $t_N$ as in \eqref{twistantitwist}, we obtain
\begin{equation} \label{twistk}
k\,=\,\kappa(-1)^{1+t_N}\frac{\left(n_N+n_S\right)^2\left(n_N^2-n_Nn_S+n_S^2\right)-9n_Nn_Sp_\mathcal{B}^2}{n_Nn_S\left(n_N-n_S\right)\left(3\left(n_N+n_S\right)^2+9p_\mathcal{B}^2\right)}\,, \quad \text{Twist}\,,
\end{equation}
and for the anti-twist class, labeled by $t_N$ as in \eqref{twistantitwist}, we have
\begin{equation} \label{atwistk}
k\,=\,\kappa(-1)^{t_N}\frac{\left(n_N-n_S\right)^2\left(n_N^2+n_Nn_S+n_S^2\right)+9n_Nn_Sp_\mathcal{B}^2}{n_Nn_S\left(n_N+n_S\right)\left(3\left(n_N-n_S\right)^2+9p_\mathcal{B}^2\right)}\,, \quad \text{Anti-Twist}\,.
\end{equation}

\subsubsection{Central charge} \label{sec333}

The 4d central charge is given by $a=\pi{R}^3/(8G_N^{(5)})$, $e.g.$, in \cite{Arav:2022lzo}, or equivalently, by $a=\pi^3N^2/(4\text{vol}_{SE_5})$, $e.g.$, in (21) of \cite{Ferrero:2020laf},{\footnote{We correct a typographical error in (21) of \cite{Ferrero:2020laf}: $\pi^2\rightarrow\pi^3$.}}  where $R$ is the $AdS_5$ radius and $\text{vol}_{SE_5}$ is the volume of Sasaki-Einstain manifolds. Thus the five-dimensional Newton's constant is given by $(G_N^{(5)})^{-1}=2\pi^2N^2/(R^3\text{vol}_{SE_5})$. From $\text{vol}_{S^5}=\pi^3$ and $\text{vol}_{T^{1,1}}=16\pi^3/27$, we can find the Newton's constant for the corresponding geometry: for $S^5$, we have $(G_N^{(5)})^{-1}=2N^2/\left(\pi{R}^3\right)$ and, for $T^{1,1}$, we find $(G_N^{(5)})^{-1}=27N^2/\left(8\pi{R}^3\right)$.

The three-dimensional Newton's constant with a unit $AdS_3$ radius is $(G_N^{(3)})^{-1}=(G_N^{(5)})^{-1}\Delta{z}\int_{y_N}^{y_S}e^V|fh|dy$ (not in conformal gauge \eqref{confgauge} here) and 2d holographic central charge is $c=(3/2)(G_N^{(3)})^{-1}$.

The integrand in the central charge integral is expressed by a total derivative,
\begin{equation}
e^Vfh\,=\,ke^{2V}f\sin\xi\,=\,-\frac{k}{2\kappa}\left(e^{3V}\cos\xi\right)'\,,
\end{equation}
and thus the central charge is determined by the pole data. As we are in conformal gauge, \eqref{confgauge}, the integrand is $-e^{2V}|h|$. Also as we have chosen $h\le0$ in \eqref{hzero}, we remove the absolute value in the integrand with a minus sign and find $e^{2V}h$. Hence, we obtain
\begin{align}
c\,=&\,81N^2\left(\frac{1}{2}\right)^3(-\frac{k}{2\kappa})\left[e^{3V}\cos\xi\right]_N^S \notag \\
=&\,-\frac{81N^2k}{16\kappa}\left(-M_{(1)}^3|_Se^{-12u_{3S}}(-1)^{t_S}+M_{(1)}^3|_Ne^{-12u_{3N}}(-1)^{t_N}\right)\,,
\end{align}
where we employed $\Delta{z}=2\pi$. Employing the expressions of $k$ and $M_{(1)}$ in terms of $n_N$, $n_S$, $t_S$ and $p_{\mathcal{B}}$ from the previous results, we find the central charges.

For the twist class, we obtain
\begin{equation} \label{ctwist}
c\,=\,\kappa(-1)^{t_N}\frac{3\left(n_S+n_N\right)\left[\left(n_N+n_S\right)^2-9p_\mathcal{B}^2\right]\left[3\left(n_N+n_S\right)^2+9p_\mathcal{B}^2\right]}{16n_Nn_S\left[\left(n_N+n_S\right)^2\left(n_N^2-n_Nn_S+n_S^2\right)-9n_Nn_Sp_\mathcal{B}^2\right]}N^2\,.
\end{equation}
By the constraints, $c>0$, $M_{(1)}|_{N,S}<0$ below \eqref{defm1m2}, and $0<e^{-12u_{3N,S}}\le1$ in \eqref{u301}, the twist class is excluded.

For the anti-twist class, we obtain
\begin{equation} \label{catwist}
c\,=\,\kappa(-1)^{t_N+1}\frac{3\left(n_N-n_S\right)\left[\left(n_N-n_S\right)^2-9p_\mathcal{B}^2\right]\left[3\left(n_N-n_S\right)^2+9p_\mathcal{B}^2\right]}{16n_Nn_S\left[\left(n_N-n_S\right)^2\left(n_N^2+n_Nn_S+n_S^2\right)+9n_Nn_Sp_\mathcal{B}^2\right]}N^2\,.
\end{equation}
To have $c>0$, we impose the following constraints,
\begin{align} \label{atwistcond}
t_N\,=\,0\,,\kappa\,>\,0\,, \quad \text{or} \quad t_N\,=\,1\,,\kappa\,<\,0\,, \qquad \Rightarrow \qquad \left(n_S-n_N\right)\,>\,3|p_\mathcal{B}|\ge0\,, \notag \\
t_N\,=\,0\,,\kappa\,<\,0\,, \quad \text{or} \quad t_N\,=\,1\,,\kappa\,>\,0\,, \qquad \Rightarrow \qquad \left(n_N-n_S\right)\,>\,3|p_\mathcal{B}|\ge0\,.
\end{align}
From the constraints listed below \eqref{spindletwist}, to have $p_0$ to be an integer, $n_N-n_S$ should be a multiplet of 3 and $p_\mathcal{B}$ is an integer. The fluxes can be expressed by
\begin{align}
p_0\,=&\,(-1)^{t_N}\frac{n_S-n_N}{3n_Nn_S}\,, \notag \\
p_1\,=&\,(-1)^{t_N}\frac{n_S-n_N}{3n_Nn_S}+p_\mathcal{B}\,, \notag \\
p_2\,=&\,(-1)^{t_N}\frac{n_S-n_N}{3n_Nn_S}-p_\mathcal{B}\,.
\end{align}
It is interesting to notice that, upon a rescaling, $3p_\mathcal{B}=2p_F$, the expression of the central charge becomes identical, up to an overall factor, to that of the Leigh-Strassler compactified on a spindle, \cite{Arav:2022lzo}. The difference in the scaling of $p_\mathcal{B}$ and $p_F$ is since $n_N(-1)^{t_S}+n_S(-1)^{t_N}$ is even for the Leigh-Strassler and is multiple of 3 for $T^{1,1}$. The overall factors are different by 2 as $a^{LS}=27/32a^{\mathcal{N}=4}$ and $a^{T^{1,1}}=27/16a^{\mathcal{N}=4}$.

For the case of no baryonic charge, $p_\mathcal{B}=0$, we find
\begin{equation} \label{cpbzero}
c\,=\,\kappa(-1)^{t_N+1}\frac{4\left(n_N-n_S\right)^3}{3n_Nn_S\left(n_N^2+n_Nn_S+n_S^2\right)}a^{KW}\,,
\end{equation}
where $a^{KW}$ is the central charge of the Klebanov-Witten theory, $a^{KW}=\pi{R}^3/(8G_N^{(5)})=27N^2R^3/64$, where the radius of $AdS_5$ is $R=1$. The corresponding solution of minimal gauged supergravity will be presented in section \ref{sec41}.

\section{Solving the BPS equations} \label{sec4}

\subsection{Analytic solutions for $p_\mathcal{B}=0$} \label{sec41}

For the case of no baryonic charge, $p_\mathcal{B}=0$, we find analytic $AdS_3\times\mathbbl{\Sigma}$ solutions in the anti-twist class. These solutions were previously obtained in five-dimensional minimal gauged supergravity in \cite{Ferrero:2020laf}. See appendix \ref{b1} for the subtruncation to five-dimensional minimal gauged supergravity. We do not employ the conformal gauge, \eqref{confgauge}, for the solutions with $p_\mathcal{B}=0$.

We find a solution of the BPS equations, \eqref{bpsequations}, \eqref{twotwocon} with \eqref{wfrel}. We set the scalar fields to vanish,
\begin{equation}
u_1\,=\,u_2\,=\,u_3\,=\,0\,,
\end{equation}
with $\theta=0$. The metric and the gauge fields are
\begin{align}
ds^2\,=&\,\frac{4y}{9}ds_{AdS_3}^2+\frac{y}{q(y)}dy^2+\frac{q(y)}{36y^2}c_0^2dz^2\,, \notag \\
A^0\,=\,A^1\,=\,A^2\,=&\,-\left[\frac{c_0\kappa}{6}\left(1-\frac{a}{y}\right)+\frac{s}{g}\right]dz\,,
\end{align}
with
\begin{equation}
\sin\xi\,=\,-\frac{\sqrt{q(y)}}{2y^{3/2}}\,, \qquad \cos\xi\,=\,\kappa\frac{3y-a}{2y^{3/2}}\,.
\end{equation}
The function, $q(y)$, is defined by
\begin{equation}
q(y)\,=\,4y^3-9y^2+6ay-a^2\,,
\end{equation}
and the constants, $a$ and $c_0$, are given by
\begin{align}
a\,=&\,\frac{\left(n_S-n_N\right)^2\left(2n_S+n_N\right)^2\left(n_S+2n_N\right)^2}{4\left(n_S^2+n_Sn_N+n_N^2\right)^3}\,, \notag \\
c_0\,=&\,\frac{2\left(n_S^2+n_Sn_N+n_N^2\right)}{2n_Sn_N\left(n_S+n_N\right)}\,.
\end{align}
The range of $y$ coordinate, $y\in[y_N,y_S]$, is given by the two smallest roots of $q(y)$,
\begin{equation}
y_N\,=\,\frac{\left(n_S^2+n_Sn_N-2n_N^2\right)^2}{4\left(n_S^2+n_Sn_N+n_N^2\right)}\,, \qquad y_S\,=\,\frac{\left(n_N^2+n_Sn_N-2n_S^2\right)^2}{4\left(n_S^2+n_Sn_N+n_N^2\right)}\,.
\end{equation}

The holographic central charge of the solution can be calculated and precisely matches \eqref{cpbzero} for $(-1)^{t_N}\kappa=+1$.

\subsection{Numerical solutions for $p_\mathcal{B}\ne0$}

In section \ref{sec33}, we have determined the values of the scalar fields, $u_2$ and $u_3$, and the metric function, $V$, in \eqref{poleu3} and \eqref{poleu2V}, in terms of the spindle data, $n_N$, $n_S$, $t_{N,S}$. We have also fixed $k$ in \eqref{atwistk}. We only have to specify the value of the scalar field, $u_1$, at the poles. Employing these results for boundary conditions, we can numerically construct $AdS_3\times\mathbbl{\Sigma}$ solutions in the anti-twist class by solving the BPS equations.

In order to solve the BPS equations numerically, we start the integration at $y=y_N$ and we choose $y_N=0$. At the poles, we have $\sin\xi=0$. We scan over the initial value of $u_1$ at $y=y_N$ in search of a solution for which we have $\sin\xi=0$ in a finite range, $i.e.$, at $y=y_S$. If we find a compact spindle solution, our boundary conditions guarantee the fluxes to be properly quantized.{\footnote {The sign of $p_\mathcal{B}$ can be switched by the symmetry, \eqref{u2a1a2}.}} We also find that \eqref{atwistcond} could be a sufficient condition to have a compact spindle solution. 

\begin{figure}[t]
\begin{center}
\includegraphics[width=3.2in]{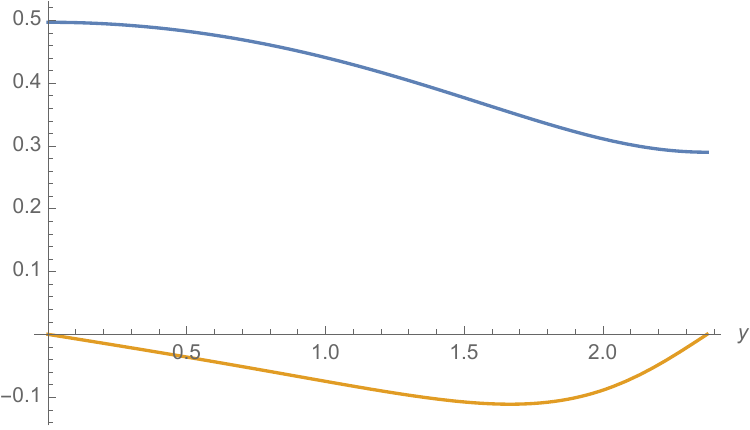} \qquad \includegraphics[width=3.2in]{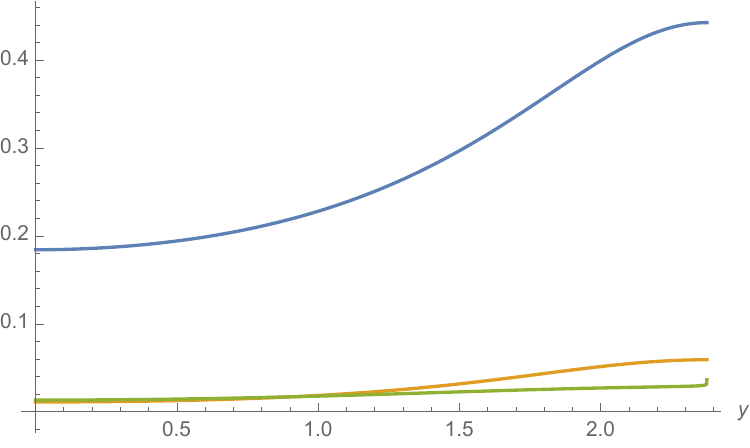}
\caption{{\it A representative $AdS_3\times\mathbbl{\Sigma}$ solution in the anti-twist class for $n_N=7$, $n_S=1$ and $p_\mathcal{B}=1$ in the range of $y=[y_N,y_S]=[0,2.375]$. The metric functions, $e^V$ (Blue) and $h$ (Orange), are on the left. The scalar fields, $u_2$ (Blue), $u_3$ (Orange), and $u_1$ (Green), are in the right. Note that $h$ vanishes at the poles.} \label{fig1}}
\end{center}
\end{figure}

We numerically perform the central charge integral in section \ref{sec333} and the result matches the analytic central charge in \eqref{catwist} with the numerical accuracy of order $10^{-6}$. We present a representative solution in figure \ref{fig1} for $n_N=7$, $n_S=1$ and $p_\mathcal{B}=1$ in the range of $y=[y_N,y_S]=[0,2.375]$. The scalar field, $u_1$, takes the values, $u_1|_N\sim0.01317$ and $u_1|_S\sim0.03648$, at the poles. Note that $h$ vanishes at the poles.

\section{Gravitational blocks} \label{sec5}

In this section, we present the gravitational block calculation for the Klebanov-Witten theory. We closely follow the analysis in section 8.1 of \cite{Boido:2022mbe}.{\footnote{We are happy to thank Jerome Gauntlett for kindly providing us with the calculation presented in this section.}}

The toric data for $T^{1,1}$ is given by four-dimensional inward-pointing normal vectors,
\begin{equation}
v_{1i}\,=\,\left(1,\vec{w}_1\right)\,, \qquad v_{2i}\,=\,\left(1,\vec{w}_2\right)\,, \qquad v_{3i}\,=\,\left(1,\vec{w}_3\right)\,, \qquad v_{4i}\,=\,\left(1,\vec{w}_4\right)\,,
\end{equation}
with
\begin{equation}
\vec{w}_1\,=\,\left(0,0\right)\,, \qquad \vec{w}_2\,=\,\left(1,0\right)\,, \qquad \vec{w}_3\,=\,\left(1,1\right)\,, \qquad \vec{w}_4\,=\,\left(0,1\right)\,.
\end{equation}
A one-dimensional kernel, $q_1^a$, satisfying $\sum_{a=1}^4q_1^av_{ai}=0$ is 
\begin{equation}
q_1^a\,=\,\left(-1,1,-1,1\right)\,,
\end{equation}
that gives the baryonic direction.

For the trivial $R$-charges of the fields, $\Delta_a$, associated with the toric divisors of $C(T^{1,1})$,
\begin{equation} \label{defdela}
\sum_{a=1}^4\Delta_a\,=\,2\,,
\end{equation}
the superpotential of the theory has R-charge 2. For the background magnetic fluxes for $U(1)^4$ symmetry with the field strengths, $F_a$, $a=1,\ldots,4$, give
\begin{equation} 
\mathfrak{p}_a\,=\,\frac{1}{2\pi}\int_{\mathbbl{\Sigma}}F_a\,,
\end{equation}
where $\mathfrak{p}_a=k_a/(n_Sn_N)$, $k_a\in\mathbb{Z}$. This implies twisting the associated fields into sections of $\mathcal{O}(\mathfrak{p}_a)$ over the spindle and the constraint to preserve supersymmetry is
\begin{equation}
\sum_{a=1}^4\mathfrak{p}_a\,=\,-\frac{\sigma{n}_N+n_S}{n_Nn_S}\,,
\end{equation}
for the twist, $\sigma=+1$, and anti-twist, $\sigma=-1$.

We introduce quantities,
\begin{align} \label{defdelpm}
\Delta_a^+\,\equiv\,\Delta_a+\frac{1}{2}\left(\mathfrak{p}_a-\frac{r_a}{2}\frac{n_S-\sigma{n}_N}{n_Sn_N}\right)\,, \notag \\
\Delta_a^-\,\equiv\,\Delta_a-\frac{1}{2}\left(\mathfrak{p}_a+\frac{r_a}{2}\frac{n_S-\sigma{n}_N}{n_Sn_N}\right)\,,
\end{align}
where $r_a$ is a set of arbitrary variables satisfying
\begin{equation}
\sum_{a=1}^4r_a\,=\,2\,.
\end{equation}
Different choice of $r_a$ corresponds to different gauge choices and the final result will be independent of it. From \eqref{defdela}-\eqref{defdelpm} we find
\begin{equation}
\sum_{a=1}^4\Delta_a^+\,=\,2-\frac{\varepsilon}{n_N}\,, \qquad \sum_{a=1}^4\Delta_a^-\,=\,2+\frac{\varepsilon}{\sigma{n}_S}\,.
\end{equation}

The off-shell trial central charge for $c$-extremization, \cite{Benini:2012cz, Benini:2013cda}, in the large $N$ limit is, \cite{Hosseini:2021fge},
\begin{equation}
c_{\text{trial}}\,=\,\frac{3}{\varepsilon}\left(\sum_{a<b<c}\left(\vec{v}_a,\vec{v}_b,\vec{v}_c\right)\Delta_a^+\Delta_b^+\Delta_c^+-\sum_{a<b<c}\left(\vec{v}_a,\vec{v}_b,\vec{v}_c\right)\Delta_a^-\Delta_b^-\Delta_c^-\right)N^2\,,
\end{equation}
where $\left(\vec{v}_a,\vec{v}_b,\vec{v}_c\right)\equiv\text{det}\left(\vec{v}_a,\vec{v}_b,\vec{v}_c\right)$. Associated with the one baryonic direction, we first impose a partial extremization condition,
\begin{equation}
\sum_{a=1}^4q_1^a\frac{\partial{c}_\text{trial}}{\partial\Delta_a}\,=\,0\,,
\end{equation}
where $q_1^a$ specifies the embedding, $U(1)_\mathcal{B}^1\subset{U}(1)^4$. After solving this and from \eqref{defdela}, an off-shell function, $c_\text{trial}|_\text{baryonic}$, is obtained which depends on, $e.g.$, $\Delta_1$, $\Delta_2$, $\varepsilon$, and $\Delta_3$ and $\Delta_4$ have been eliminated. Then extremizing over the baryonic directions, $\Delta_a^\pm|_\text{baryonic}$ obtained are also functions of trial $R$-charges, $e.g.$, $\Delta_1$, $\Delta_2$, $\varepsilon$. After further extremizing over, $e.g.$, $\Delta_1$, $\Delta_2$, $\varepsilon$, the on-shell result is obtained. The magnetic fluxes, $\mathfrak{p}_a$, are fixed through this procedure.

We restrict to fluxes preserving $SU(2)\times{S}U(2)$ symmetry,
\begin{equation}
\mathfrak{p}_2\,=\,\mathfrak{p}_1\,, \qquad \mathfrak{p}_4\,=\,\mathfrak{p}_3\,,
\end{equation}
where $2\left(\mathfrak{p}_1+\mathfrak{p}_2\right)=-\frac{\sigma{n}_N+n_S}{n_Nn_S}$ and
\begin{equation}
\mathfrak{p}_\mathcal{B}\,\equiv\,\frac{2}{3}\left(\mathfrak{p}_1-\mathfrak{p}_3\right)\,,
\end{equation}
with the factor, $2/3$, introduced for convenience. Then on-shell we find
\begin{equation}
c\,=\,\frac{3\left(n_S+\sigma{n}_N\right)\left[\left(n_S+\sigma{n}_N\right)^2-9p_\mathcal{B}^2\right]\left[3\left(n_S+\sigma{n}_N\right)^2+9p_\mathcal{B}^2\right]}{16n_Nn_S\left[\left(n_S+\sigma{n}_N\right)^2\left(n_N^2-\sigma{n}_Nn_S+n_S^2\right)-9\sigma{n}_Nn_Sp_\mathcal{B}^2\right]}N^2\,,
\end{equation} 
where we have identified $\mathfrak{p}_\mathcal{B}=p_\mathcal{B}/\left(n_Nn_S\right)$. This result precisely matches the holographic central charge from supergravity calculation, \eqref{ctwist} and \eqref{catwist}. We also find
\begin{equation}
\varepsilon_*\,=\,\frac{n_Nn_S\left(n_S-\sigma{n}_N\right)\left(3\left(n_S+\sigma{n}_N\right)^2+9p_\mathcal{B}^2\right)}{\left(n_S+\sigma{n}_N\right)^2\left(n_N^2-\sigma{n}_Nn_S+n_S^2\right)-9\sigma{n}_Nn_Sp_\mathcal{B}^2}\,,
\end{equation}
and it precisely matches $-1/k$ from \eqref{twistk} and \eqref{atwistk}.

Although the gravitational blocks enable us to calculate the $c$-function without the knowledge of explicit solutions in supergravity, the central charge from this calculation does not guarantee the existence of the corresponding solution. Thus our solution provides an explicit test of the gravitational block calculation for the spindle solutions with baryonic charges presented in \cite{Boido:2022mbe}.

\section{Conclusions} \label{sec6}

In this paper, we have constructed the $AdS_3\times\mathbbl{\Sigma}$ solutions with baryonic charge in the Betti-vector truncation of five-dimensional gauged $\mathcal{N}=4$ supergravity, \cite{Halmagyi:2011yd}, where $\mathbbl{\Sigma}$ is a spindle. The truncation is obtained from type IIB supergravity on $AdS_5\times{T}^{1,1}$, \cite{Cassani:2010na, Bena:2010pr}. The solutions realize supersymmetry by the anti-twist. The dual field theories are 2d $\mathcal{N}=(0,2)$ SCFTs constructed from the Klebanov-Witten theory, \cite{Klebanov:1998hh}, compactified on a spindle. We have calculated the holographic central charge of the solutions and it precisely matched the result from the gravitational block calculation.

The natural question for spindle solutions discovered so far would be to reproduce their $c$-functions, $e.g.$, central charge, the Bekenstein-Hawking entropy, and free energy, from the localization calculation in dual field theory. A very recent effort, \cite{Inglese:2023wky}, would shed light on this question.

As it was noted in the main body of the paper, upon a suitable rescaling of baryonic charge number, $p_\mathcal{B}$, the expression of the central charge becomes identical to the one for the Leigh-Strassler theory compactified on a spindle, \cite{Arav:2022lzo}, up to an overall factor. Although both of them reduce to the universal solution, \cite{Ferrero:2020laf}, when the flavor charges vanish, it is interesting that they deform from the central charge of the universal solution uniformly. The Leigh-Strassler and the Klebanov-Witten theories are very distinct from the aspect of field theory and also from the uplifted solution in string theory. Our solution will be uplifted to $AdS_3\times\mathbbl{\Sigma}\times{T}^{1,1}$ solution only with non-trivial five-form flux in type IIB supergravity and is an example of the GK geometry, \cite{Kim:2005ez, Gauntlett:2007ts}. On the other hand, when uplifted, the Leigh-Strassler solution will involve other non-trivial fluxes in type IIB supergravity. Thus it would be interesting to understand how they share identical expressions of the central charge despite the very different nature of the solutions.

It would be interesting to obtain the central charge from the anomaly polynomial as it was done for the Leigh-Strassler solution, \cite{Arav:2022lzo}. Otherwise, the anomaly inflow method for type IIB supergravity developed in \cite{Bah:2020jas} might be useful. On the other hand, the gravitational block calculation is available for the GK geometries at the moment and the Leigh-Strassler solution is not in the class of GK geometry.

\bigskip
\bigskip
\leftline{\bf Acknowledgements}
\noindent We would like to thank Jerome Gauntlett for the instructive conversation and comments. This work was supported by the Kumoh National Institute of Technology. This work was also supported by the Kavli Institute for Theoretical Sciences (KITS), the University of Chinese Academy of Sciences (UCAS), and the Fundamental Research Funds for the Central Universities.

\appendix
\section{The Betti-vector truncation} \label{appA}
\renewcommand{\theequation}{A.\arabic{equation}}
\setcounter{equation}{0} 

A consistent truncation of type IIB supergravity on $T^{1,1}$ leads to gauged $\mathcal{N}=4$ supergravity in five dimensions, \cite{Cassani:2010na, Bena:2010pr}. In \cite{Halmagyi:2011yd} several subtruncations to gauged $\mathcal{N}=2$ supergravity coupled to matter multiplets were studied: the Betti-vector truncation coupled to two vector multiplets and two hyperultiplets and the Betti-hyper truncation coupled to a vector multiplet and three hypermultiplets. We will consider the Betti-vector truncation.

In the $\mathcal{N}=2$ formalism, the Betti-vector truncation consists of the graviton multiplet, two vector multiplets, and two hypermultiplets, \cite{Halmagyi:2011yd, Amariti:2016mnz}. One of the hypermultiplets is the universal hypermultiplet. This truncation admits an $\mathcal{N}=2$ supersymmetric $AdS_5$ vacuum. See \cite{Louis:2016msm} for the $AdS_5$ vacuum structure of the theory.
\begin{itemize}
\item Gravity multiplet + 2 vector multiplets: $\left(g_{\mu\nu},A^0,A^1,A^2,u_2,u_3\right)$\,,
\item 2 hypermultiplets: $\left(u_1,\theta,\tau, \bar{\tau},b_0^i, \bar{b}_0^i\right)$\,,
\end{itemize}
where $i=1,2$. In this truncation, one of the vectors can become massive by the Higgs mechanism and the vector multiplet and a hypermultiplet become a massive vector multiplet. Then the field content is
\begin{itemize}
\item Gravity multiplet: graviton, $g_{\mu\nu}$, and a massless vector, $A_g$\,,
\item A massless Betti multiplet: a massless vector, $A_\mathcal{B}$, and a real scalar, $u_2$, with $m_{u_2}^2=-4$ ($\Delta=2$)\,,
\item A universal hypermultiplet: 4 real scalars, $(\tau, \bar{\tau}, b_0^1, \bar{b}_0^1)$, with $m_\tau^2=0$ ($\Delta=4$) and $m_{b_0}^2=-3$ ($\Delta=3$)\,,
\item A massive vector multiplet: a massive vector, $A_m$, which is eaten by axion, $\theta$, with $m^2=24$ ($\Delta=7$), and 4 real scalars, ($u_3$, $u_1$, $b_0^2$, $\bar{b}_0^2$), with $m^2=12,21,21,32$ ($\Delta=6,7,7,8$)\,.
\end{itemize}
In the Betti-vector truncation, $AdS_3\times\Sigma_{\mathfrak{g}}$ solutions were constructed  in \cite{Amariti:2016mnz} where $\Sigma_{\mathfrak{g}}$ is a Riemann surface with genus, $\mathfrak{g}$. See \cite{Benini:2015bwz} also for the study on the more general case of $Y^{p,q}$ manifolds.{\footnote{We correct a few typographical errors in the BPS equations in (5.14) of \cite{Amariti:2016mnz}: $8\rightarrow4$, $24\rightarrow12$ and $e^f$ is missing in the second term of $u_1$ equation. Also, in the expression of $K_0$ in (5.1), $-2\partial_k\rightarrow-Q\partial_k$ where $Q=4$.}}

Let us consider gauged $\mathcal{N}=2$ supergravity coupled to $n_v$ vector multiplets and $n_h$ hypermultiplets in five dimensions, \cite{Ceresole:2000jd}. The gravity multiplet consists of the metric, $g_{\mu\nu}$, and the graviphoton, $A^0$. The vector multiplets have $n_v$ vectors, $A_\mu^I$, $I=1,\ldots,n_v$, and scalar fields, $\phi^x$, living on a very special manifold. The hypermultiplets have $4n_h$ scalar fields, $q^X$, living on a quaternionic manifold. The supersymmetry variations of gravitino, gaugino, and hyperino are given by, respectively,
\begin{align} \label{n2susy}
\delta\psi_{\mu{i}}\,=&\,D_\mu\epsilon_i+\frac{i}{6}X^IP_I^r\gamma_\mu\left(i\sigma^r\right)_i\,^j\epsilon_j+\frac{i}{24}X_IF^I_{\mu\nu}\left(\gamma_\mu\,^{\nu\rho}-4\delta_\mu^\nu\gamma^\rho\right)\epsilon_i\,, \notag \\
\delta\lambda_i^x\,=&\,-\frac{i}{2}D_\mu\phi^x\gamma^\mu\epsilon_i-g^{xy}\partial_yX^IP_I^r\left(i\sigma^r\right)_i\,^j\epsilon_j-\frac{1}{4}g^{xy}\partial_yX_IF^I_{\mu\nu}\gamma^{\mu\nu}\epsilon_i\,, \notag \\
\delta\zeta^A\,=&\,f_X^{iA}\left[-\frac{i}{2}D_\mu{q}^X\gamma^\mu\epsilon_i+\frac{1}{2}X^IK_I^X\epsilon_i\right]\,,
\end{align}
where we have
\begin{equation}
D_\mu\epsilon_i\,=\,\nabla_\mu\epsilon_i+\frac{1}{2}A^I_\mu{P}^r_I\left(i\sigma^r\right)_i\,^j\epsilon_j+\frac{1}{2}\partial_\mu{q}^X\omega_X^r\left(i\sigma^r\right)_i\,^j\epsilon_j\,,
\end{equation}
and $\sigma^r$, $r=1,2,3$, are the Pauli matrices. The vielbeins and connections on the quaternionic manifold are $f_i^{AX}$ and $\omega_X^r$, respectively, where $i=1,2$ and $A=1,\ldots2n_h$ are $SU(2)$ and $Sp(2n_h)$ indices. The covariant derivatives on the scalar fields are defined by
\begin{equation}
D_\mu\phi^x\,=\,\partial_\mu\phi^x+A^I_\mu{K}_I^x(\phi^x)\,, \qquad D_\mu{q}^X\,=\,\partial_\mu{q}^X+A^I_\mu{K}_I^X(q^X)\,,
\end{equation}
where $K_I^x(\phi^x)$ and $K_I^X(q^X)$ are the Killing vectors which correspond to the gauging of the isometries of the very special and quaternionic manifolds, respectively.

We will consider a subtruncation of the Betti-vector truncation with the graviton, three vector fields, $A^0$, $A^1$, $A^2$, two real scalar fields, $u_2$, $u_3$, and a complex scalar field, $(u_1,\theta)$, \cite{Halmagyi:2011yd, Amariti:2016mnz}.{\footnote{The scalar field, $\theta$, is denoted by $k$ in \cite{Halmagyi:2011yd, Amariti:2016mnz}. We also fix the parameter, $Q$, in \cite{Halmagyi:2011yd, Amariti:2016mnz} to be $Q=4$.}} For the subtruncation we consider, the scalar fields are parametrized by
\begin{align}
\phi^x\,:&\,\left(u_2,u_3\right)\,\equiv\,\left(\phi^2,\phi^3\right)\,, \notag \\
q^X\,:&\,\,\,\,\left(u_1,\theta\right)\,\equiv\,\left(q^1,q^\theta\right)\,.
\end{align}
The non-trivial components of the Killing prepotential, $P_I^r$, the Killing vectors on the quaternionic manifold, $K_I^X$, and the scalar fields, $X^I$, are parametrized by
\begin{align}
P_I^3\,=&\,\left(3-2e^{-4u_1},e^{-4u_1},e^{-4u_1}\right)\,, \notag \\
K_I^\theta\,=&\,\left(-4,2,2\right)\,, \notag \\
X^I\,=&\,\left(e^{4u_3},e^{2u_2-2u_3},e^{-2u_2-2u_3}\right)\,,
\end{align}
and $X_I\,=\,\left(X^I\right)^{-1}$. We define the quantities including the superpotential, $W$,
\begin{align}
W\,\equiv&\,X^IP_I^3\,, \notag \\
H_{\mu\nu}\,\equiv&\,-\frac{1}{2}X_IF^I_{\mu\nu}\,, \notag \\
Q_\mu\,\equiv&\,-\frac{1}{2}A_\mu^IP_I^3-\frac{1}{2}\partial_\mu{q}^X\omega_X^3\,.
\end{align}
The supersymmetry variations of gravitino, gaugino, and hyperino, \eqref{n2susy}, reduce to
\begin{align}
&6iD_\mu\epsilon_i-W\gamma_\mu\left(i\sigma^3\right)_i\,^j\epsilon_j+\frac{1}{2}H_{\nu\rho}\left(\gamma_\mu\,^{\nu\rho}-4\delta_\mu^\nu\gamma^\rho\right)\epsilon_i\,=\,0, \notag \\
&8i\partial_\mu{u}_1\gamma^\mu\epsilon_i+\partial_{u_1}W\left(i\sigma^3\right)_i\,^j\epsilon_j+2\partial_{u_1}Q_\mu\gamma^\mu\epsilon_i\,=\,0\,, \notag \\
&4i\partial_\mu{u}_2\gamma^\mu\epsilon_i+\partial_{u_2}W\left(i\sigma^3\right)_i\,^j\epsilon_j-\frac{1}{2}\partial_{u_2}H_{\mu\nu}\gamma^{\mu\nu}\epsilon_i\,=\,0\,, \notag \\
&12i\partial_\mu{u}_3\gamma^\mu\epsilon_i+\partial_{u_3}W\left(i\sigma^3\right)_i\,^j\epsilon_j-\frac{1}{2}\partial_{u_3}H_{\mu\nu}\gamma^{\mu\nu}\epsilon_i\,=\,0\,,
\end{align}
where we have 
\begin{equation}
D_\mu\epsilon_i\,=\,\nabla_\mu\epsilon_i-Q_\mu\left(i\sigma^3\right)_i\,^j\epsilon_j\,,
\end{equation}
with
\begin{align} \label{whqdef}
W\,=&\,\left(e^{2u_2-2u_3}+e^{-2u_2-2u_3}-2e^{4u_3}\right)e^{-4u_1}+3e^{4u_3}\,, \notag \\
H_{\mu\nu}\,=&\,-\frac{1}{2}\left(e^{-4u_3}F^0_{\mu\nu}+e^{-2u_2+2u_3}F^1_{\mu\nu}+e^{2u_2+2u_3}F^2_{\mu\nu}\right)\,, \notag \\
Q_\mu\,=&\,-\frac{1}{2}\Big[\left(3-2e^{-4u_1}\right)A^0_\mu+e^{-4u_1}\left(A^1_\mu+A^2_\mu\right)\Big]+\frac{1}{2}e^{-4u_1}D_\mu\theta\,.
\end{align}
We introduce the supersymmetry parameter, $\epsilon\equiv\epsilon_1+\epsilon_2$, and the supersymmetry variations can be rewritten by
\begin{align} \label{susyapp}
\left[6\nabla_\mu-6iQ_\mu-W\gamma_\mu-\frac{i}{2}H_{\nu\rho}\left(\gamma_\mu\,^{\nu\rho}-4\delta_\mu^\nu\gamma^\rho\right)\right]\epsilon\,=\,0\,, \notag \\
\Big[8\partial_\mu{u}_1\gamma^\mu+\partial_{u_1}W+2i\partial_{u_1}Q_\mu\gamma^\mu\Big]\epsilon\,=\,0\,, \notag \\
\left[4\partial_\mu{u}_2\gamma^\mu+\partial_{u_2}W+\frac{i}{2}\partial_{u_2}H_{\mu\nu}\gamma^{\mu\nu}\right]\epsilon\,=\,0\,, \notag \\
\left[12\partial_\mu{u}_3\gamma^\mu+\partial_{u_3}W+\frac{i}{2}\partial_{u_3}H_{\mu\nu}\gamma^{\mu\nu}\right]\epsilon\,=\,0\,.
\end{align}

The bosonic Lagrangian of the subtruncation in a mostly plus signature is given by, \cite{Halmagyi:2011yd},
\begin{align} \label{bvlag2}
e^{-1}\mathcal{L}\,=\,&R-8\partial_\mu{u}_1\partial^\mu{u}_1-\frac{1}{2}e^{-8u_1}D_\mu\theta{D}^\mu\theta-4\partial_\mu{u}_2\partial^\mu{u}_2-12\partial_\mu{u}_3\partial^\mu{u}_3-V \notag \\
&-\frac{1}{4}e^{-8u_3}F^0_{\mu\nu}F^{0\mu\nu}-\frac{1}{4}e^{-4u_2+4u_3}F^1_{\mu\nu}F^{1\mu\nu}-\frac{1}{4}e^{4u_2+4u_3}F^2_{\mu\nu}F^{2\mu\nu}+\frac{1}{4}\epsilon^{\mu\nu\rho\sigma\delta}F^1_{\mu\nu}F^2_{\rho\sigma}A^0_\delta\,,
\end{align}
where we define
\begin{equation} \label{dtheta}
D\theta\,\equiv\,d\theta-4A^0+2A^1+2A^2\,.
\end{equation}
The scalar potential is
\begin{equation}
V\,=\,\frac{1}{8}\left(\frac{\partial{W}}{\partial{u_1}}\right)^2+\frac{1}{4}\left(\frac{\partial{W}}{\partial{u_2}}\right)^2+\frac{1}{12}\left(\frac{\partial{W}}{\partial{u_3}}\right)^2-\frac{4}{3}W^2\,.
\end{equation}
The Betti vector, $A_{\mathcal{B}}$, the massless vector appearing in the Sasaki-Einstein truncation, $A_g$, and the massive vector, $A_m$, are given by, \cite{Amariti:2016mnz},
\begin{equation}
A_{\mathcal{B}}\,=\,\frac{A^2-A^1}{\sqrt{2}}\,, \qquad A_m\,=\,\frac{A^1+A^2-2A^0}{\sqrt{6}}\,, \qquad A_g\,=\,\frac{A^0+A^1+A^2}{\sqrt{3}}\,,
\end{equation}
and, when $A^1\,=\,A^2$, the Betti vector vanishes and after a suitable truncation of the scalar fields, it returns to the Sasaki-Einstein truncation. The $\mathcal{N}=2$ $AdS_5$ supersymmetric vacuum with unit radius is located where the scalar fields are trivial. 

We present the equations of motion from the Lagrangian, \eqref{bvlag2}. The Einstein equations are
\begin{align}
R_{\mu\nu}-\frac{1}{2}Rg_{\mu\nu}+&\frac{1}{2}Vg_{\mu\nu}-8T^{u_1}_{\mu\nu}-4T^{u_2}_{\mu\nu}-12T^{u_3}_{\mu\nu}-\frac{1}{2}e^{-8u_1}T^\theta_{\mu\nu} \notag \\
-&\frac{1}{2}\left(e^{-8u_3}T^{A^0}_{\mu\nu}+e^{-4u_2+4u_3}T^{A^1}_{\mu\nu}+e^{4u_2+4u_3}T^{A^2}_{\mu\nu}\right)\,=\,0\,,
\end{align}
where the energy-momentum tensors for a scalar field, $X$, and the gauge fields, $A^I$, are, respectively,
\begin{align}
T^X_{\mu\nu}\,=&\,\partial_\mu{X}\partial_\nu{X}-\frac{1}{2}g_{\mu\nu}\partial_\rho{X}\partial^\rho{X}\,, \notag \\
T^{A^I}_{\mu\nu}\,=&\,g^{\rho\sigma}F^I_{\mu\rho}F^I_{\nu\sigma}-\frac{1}{4}g_{\mu\nu}F^I_{\rho\sigma}F^{\rho\sigma}\,.
\end{align}
The scalar field equations are 
\begin{align}
&\frac{1}{\sqrt{-g}}\partial_\mu\left(\sqrt{-g}g^{\mu\nu}\partial_\nu{u}_1\right)-\frac{1}{16}\frac{\partial{V}}{\partial{u_1}}+\frac{1}{4}e^{-8u_1}D_\mu\theta{D}^\mu\theta\,=\,0\,, \notag \\
&\frac{1}{\sqrt{-g}}\partial_\mu\left(\sqrt{-g}g^{\mu\nu}\partial_\nu{u}_2\right)-\frac{1}{8}\frac{\partial{V}}{\partial{u_2}}+\frac{1}{8}\left(e^{-4u_2+4u_3}F^1_{\mu\nu}F^{1\mu\nu}-e^{4u_2+4u_3}F^2_{\mu\nu}F^{2\mu\nu}\right)\,=\,0\,, \notag \\
&\frac{1}{\sqrt{-g}}\partial_\mu\left(\sqrt{-g}g^{\mu\nu}\partial_\nu{u}_3\right)-\frac{1}{24}\frac{\partial{V}}{\partial{u_3}}+\frac{1}{24}\left(2e^{-8u_3}F^0_{\mu\nu}F^{0\mu\nu}-e^{-4u_2+4u_3}F^1_{\mu\nu}F^{1\mu\nu}-e^{4u_2+4u_3}F^2_{\mu\nu}F^{2\mu\nu}\right)\,=\,0\,,
\end{align}
and the Maxwell equations are
\begin{align}
\partial_\nu\left(\sqrt{-g}e^{-8u_3}F^{0\mu\nu}\right)-4\sqrt{-g}e^{-8u_1}g^{\mu\nu}D_\nu\theta\,=\,0\,, \notag \\
\partial_\nu\left(\sqrt{-g}e^{-4u_2+4u_3}F^{1\mu\nu}\right)+2\sqrt{-g}e^{-8u_1}g^{\mu\nu}D_\nu\theta\,=\,0\,, \notag \\
\partial_\nu\left(\sqrt{-g}e^{4u_2+4u_3}F^{2\mu\nu}\right)+2\sqrt{-g}e^{-8u_1}g^{\mu\nu}D_\nu\theta\,=\,0\,.
\end{align}

\subsection{Truncation to minimal gauged supergravity via KW} \label{b1}

The truncation to minimal gauged supergravity associated to the Klebanov-Witten fixed point is obtained by setting
\begin{equation}
u_1\,=\,u_2\,=\,u_3\,=\,\theta\,=\,0\,, \qquad A\,\equiv\,A^0\,=\,A^1\,=\,A^2\,.
\end{equation}
The Lagrangian in \eqref{bvlag2} reduces to
\begin{equation}
e^{-1}\mathcal{L}\,=\,R+12-\frac{3}{4}F_{\mu\nu}F^{\mu\nu}+\frac{1}{4}\epsilon^{\mu\nu\rho\sigma\delta}F_{\mu\nu}F_{\rho\sigma}A_\delta\,,
\end{equation}
where $F=dA$. If we redefine $F\rightarrow\frac{2}{3}F$, we obtain the Lagrangian in the normalization of \cite{Ferrero:2020laf}.

\section{Supersymmetry variations} \label{appB}
\renewcommand{\theequation}{B.\arabic{equation}}
\setcounter{equation}{0} 

\subsection{Derivation of the BPS equations}

Following the analysis in \cite{Arav:2022lzo}, we derive the BPS equations from the supersymmetry variations in \eqref{susyapp} for the $AdS_3$ ansatz in \eqref{ansatzads3}.

We employ the gamma matrices of mostly plus signature,
\begin{equation}
\gamma^m\,=\,\Gamma^m\otimes\sigma^3\,, \qquad \gamma^3\,=\,1\otimes\sigma^1\,, \qquad \gamma^4\,=\,1\otimes\sigma^2\,,
\end{equation}
where gamma matrices in three dimensions are $\Gamma^m=\left(i\sigma^2,\sigma^3,\sigma^1\right)$.
With the orthonormal basis in \eqref{orthobasis}, we employ the Killing spinors,
\begin{equation}
\epsilon\,=\,\psi\otimes\chi\,,
\end{equation}
where $\psi$ is a two-component spinor on $AdS_3$,
\begin{equation}
D_m\psi\,=\,\frac{1}{2}\kappa\Gamma_m\psi\,,
\end{equation}
and $\kappa=\pm1$.

From the gravitino variation, \eqref{susyapp}, in the coordinates tangent to $AdS_3$, we find
\begin{equation}
\left[-i\left(3\kappa{e}^{-V}+H_{34}\right)\gamma^{34}+3V'f^{-1}\gamma^3\right]\,=\,W\epsilon\,.
\end{equation}
The left-hand side should have eigenvalue, $W$, and we find a projection condition,
\begin{equation}
\left[i\cos\xi\gamma^{34}+\sin\xi\gamma^3\right]\epsilon\,=\,+\epsilon\,,
\end{equation}
with
\begin{equation} \label{cossin}
-3\kappa{e}^{-V}-H_{34}\,=\,W\cos\xi\,, \qquad 3V'f^{-1}\,=\,W\sin\xi\,.
\end{equation}
The projection condition has a solution,
\begin{equation}
\epsilon\,=\,e^{i\frac{\xi}{2}\gamma^4}\eta\,, \qquad \gamma^3\eta\,=\,+i\gamma^4\eta\,.
\end{equation}
From \eqref{cossin} we find $\partial_z\xi=0$. At $\xi=0,\pi$, the spinors have a definite chirality with respect to $\gamma^{34}$,
\begin{equation}
\xi\,=\,0,\pi\,, \qquad \gamma^{34}\epsilon\,=\,\pm{i}\epsilon\,.
\end{equation}

From the gravitino variation, \eqref{susyapp}, in the $y$ coordinate, by employing \eqref{cossin}, we find
\begin{equation} \label{a1a21}
\left[\partial_y-\frac{1}{2}V'+\frac{i}{2}\left(\partial_y\xi+fH_{34}+\kappa{f}e^{-V}\right)\gamma^4\right]\eta\,=\,0\,.
\end{equation}
From the $z$ coordinate, we obtain
\begin{align} \label{a1a22}
\Big[\partial_z-iQ_z&+\frac{i}{2}f^{-1}h'\cos\xi-\frac{i}{3}H_{34}h\sin\xi \notag \\
&-\left(-\frac{1}{2}f^{-1}h'\sin\xi+\frac{Wh}{6}-\frac{1}{3}H_{34}h\cos\xi\right)\gamma^4\Big]\eta\,=\,0\,.
\end{align}

An expression, $\left(a_1+ia_2\gamma^4\right)\eta=0$, requires $a_1^2+a_2^2=0$. Thus we find from \eqref{a1a21} and \eqref{a1a22},
\begin{equation}
\eta\,=\,e^{V/2}e^{isz}\eta_0\,,
\end{equation}
where $\eta_0$ is independent of $y$ and $z$ and 
\begin{align}
\partial_y\xi+fH_{34}+\kappa{f}e^{-V}\,=&\,0\,, \notag \\
\left(s-Q_z\right)+\frac{1}{2}f^{-1}h'\cos\xi-\frac{1}{3}H_{34}h\sin\xi\,=&\,0\,, \notag \\
-\frac{1}{2}f^{-1}h'\sin\xi+\frac{Wh}{6}-\frac{1}{3}H_{34}h\cos\xi\,=&\,0\,.
\end{align}
From these we obtain
\begin{align}
f^{-1}h'\,=&\,\frac{Wh}{3}\sin\xi-2\left(s-Q_z\right)\cos\xi\,, \notag \\
hH_{34}\,=&\,\frac{Wh}{2}\cos\xi+3\left(s-Q_z\right)\sin\xi\,,
\end{align}
and, with the first relation in \eqref{cossin}, we obtain
\begin{equation}
\left(s-Q_z\right)\sin\xi\,=\,-\frac{1}{2}Wh\cos\xi-\kappa{h}e^{-V}\,,
\end{equation}
and thus we have
\begin{align}
H_{34}\,=&\,-W\cos\xi-3\kappa{e}^{-V}\,, \notag \\
f^{-1}\partial_y\xi\,=&\,W\cos\xi+2\kappa{e}^{-V}\,.
\end{align}
We solve for $\left(s-Q_z\right)$ for $\sin\xi\ne0$ and find
\begin{equation}
f^{-1}\frac{h'}{h}\sin\xi\,=\,2\kappa{e}^{-V}\cos\xi+\frac{W}{3}\left(1+2\cos^2\xi\right)\,.
\end{equation}

In the identical procedure, from the gaugino variations, we find
\begin{align}
f^{-1}u_2'+\frac{1}{4}\partial_{u_2}W\sin\xi\,=&\,0\,, \notag \\
f^{-1}u_3'+\frac{1}{12}\partial_{u_3}W\sin\xi\,=&\,0\,,
\end{align}
with
\begin{align}
\partial_{u_2}W\cos\xi+\partial_{u_2}H_{34}\,=\,0\,, \notag \\
\partial_{u_3}W\cos\xi+\partial_{u_3}H_{34}\,=\,0\,.
\end{align}
We find expressions of the field strengths of the gauge fields given by
\begin{align} 
e^{-4u_3}F^0_{34}\,=&\,\frac{1}{3}\left(2W+\partial_{u_3}W\right)\cos\xi+2\kappa{e}^{-V}\,, \notag \\
e^{-2u_2+2u_3}F^1_{34}\,=&\,\frac{1}{6}\left(4W+3\partial_{u_2}W-\partial_{u_3}W\right)\cos\xi+2\kappa{e}^{-V}\,, \notag \\
e^{2u_2+2u_3}F^2_{34}\,=&\,\frac{1}{6}\left(4W-3\partial_{u_2}W-\partial_{u_3}W\right)\cos\xi+2\kappa{e}^{-V}\,.
\end{align}

From the hyperino variation, we find
\begin{align}
f^{-1}u_1'+\frac{1}{8}\partial_{u_1}W\sin\xi+\frac{1}{4}\partial_{u_1}Q_z\cos\xi{h}^{-1}\,=&\,0\,, \notag \\
\frac{1}{2}\partial_{u_1}W\cos\xi-\partial_{u_1}Q_z\sin\xi{h}^{-1}\,=&\,0\,.
\end{align}

{\bf Summary:} Assuming $\sin\xi\ne0$ the BPS equation are given by
\begin{align} 
f^{-1}\xi'\,=&\,W\cos\xi+2\kappa{e}^{-V}\,, \notag \\
f^{-1}V'\,=&\,\frac{1}{3}W\sin\xi\,, \notag \\
f^{-1}u_1'\,=&\,-\frac{1}{8}\frac{\partial_{u_1}W}{\sin\xi}\,, \notag \\
f^{-1}u_2'\,=&\,-\frac{1}{4}\partial_{u_2}W\sin\xi\,, \notag \\
f^{-1}u_3'\,=&\,-\frac{1}{12}\partial_{u_3}W\sin\xi\,, \notag \\
f^{-1}\frac{h'}{h}\sin\xi\,=&\,2\kappa{e}^{-V}\cos\xi+\frac{W}{3}\left(1+2\cos^2\xi\right)\,,
\end{align}
with two constraints,
\begin{align} 
\left(s-Q_z\right)\sin\xi\,=&\,-\frac{1}{2}Wh\sin\xi-\kappa{h}e^{-V}\,, \notag \\
\frac{1}{2}\partial_{u_1}\cos\xi\,=&\,\partial_{u_1}Q_z\sin\xi{h}^{-1}\,.
\end{align}
The field strengths of the gauge fields are given by
\begin{align} 
e^{-4u_3}F^0_{34}\,=&\,\frac{1}{3}\left(2W+\partial_{u_3}W\right)\cos\xi+2\kappa{e}^{-V}\,, \notag \\
e^{-2u_2+2u_3}F^1_{34}\,=&\,\frac{1}{6}\left(4W+3\partial_{u_2}W-\partial_{u_3}W\right)\cos\xi+2\kappa{e}^{-V}\,, \notag \\
e^{2u_2+2u_3}F^2_{34}\,=&\,\frac{1}{6}\left(4W-3\partial_{u_2}W-\partial_{u_3}W\right)\cos\xi+2\kappa{e}^{-V}\,, \notag \\
H_{34}\,=&\,-W\cos\xi-3\kappa{e}^{-V}\,.
\end{align}

Along the BPS flow, we find
\begin{equation}
\partial_yW\,=\,-gf\sin\xi\left[\frac{1}{4}\left(\partial_{u_2}W\right)^2+\frac{1}{12}\left(\partial_{u_3}W\right)^2+\frac{1}{8\sin^2\xi}\left(\partial_{u_1}W\right)^2\right]\,,
\end{equation}
and, if the sign of $f\sin\xi$ does not change, $W$ is monotonic along the flow.

There is an integral of the BPS equations,
\begin{equation}
he^{-V}\,=\,k\sin\xi\,,
\end{equation}
where $k$ is a constant. By employing the integral of the BPS equations to eliminate $h$, the BPS equations are given by
\begin{align} 
f^{-1}\xi'\,=&\,-2k^{-1}\left(s-Q_z\right)e^{-V}\,, \notag \\
f^{-1}V'\,=&\,\frac{1}{3}W\sin\xi\,, \notag \\
f^{-1}u_1'\,=&\,-\frac{1}{8}\frac{\partial_{u_1}W}{\sin\xi}\,, \notag \\
f^{-1}u_2'\,=&\,-\frac{1}{4}\partial_{u_2}W\sin\xi\,, \notag \\
f^{-1}u_3'\,=&\,-\frac{1}{12}\partial_{u_3}W\sin\xi\,,
\end{align}
with two constraints,
\begin{align} \label{twotwotwo}
\left(s-Q_z\right)\sin\xi\,=&\,-k\left(\frac{1}{2}We^V\cos\xi+\kappa\right)\,, \notag \\
\frac{1}{2}\partial_{u_1}\cos\xi\,=&\,k^{-1}e^{-V}\partial_{u_1}Q_z\,.
\end{align}

From $Q_\mu$ in \eqref{whqdef}, we obtain
\begin{equation}
\partial_{u_1}Q_z\,=\,-2e^{-4u_1}D_z\theta-2e^{-4u_1}\left(2A^0-A^1-A^2\right)\,.
\end{equation}
With the second constraint in \eqref{twotwotwo}, we find
\begin{equation}
D_z\theta\,=\,\frac{ke^V\partial_{u_1}W\cos\xi}{4e^{-4u_1}}-\left(2A_z^0-A_z^1-A_z^2\right)\,.
\end{equation}
From this constraint, we find that the field strengths are expressed by
\begin{equation}
F^I_{yz}\,=\,\left(a^I\right)'\,=\,\left(\mathcal{I}^I\right)'\,,
\end{equation}
where we define
\begin{align} \label{introi1i2i3}
\mathcal{I}^{(0)}\,\equiv&\,-ke^V\cos\xi{e}^{4u_3}\,, \notag \\
\mathcal{I}^{(1)}\,\equiv&\,-ke^V\cos\xi{e}^{2u_2-2u_3}\,, \notag \\ 
\mathcal{I}^{(2)}\,\equiv&\,-ke^V\cos\xi{e}^{-2u_2-2u_3}\,.
\end{align}

There is a symmetry of the BPS equations,
\begin{equation} \label{hsymm}
h\,\rightarrow\,-h\,, \qquad z\,\rightarrow\,-z\,,
\end{equation}
when we have $Q_z\rightarrow-Q_z$, $s\rightarrow-s$, $a^I\rightarrow-a^I$, $k\rightarrow-k$ and $F_{34}^I\rightarrow+F_{34}^I$. The frame is invariant under this transformation. We fix $h\le0$ by this symmetry in the main text. There is also a $\mathbb{Z}_2$ symmetry of the model,
\begin{equation}
u_2\,\rightarrow\,-u_2\,, \qquad A^1\,\leftrightarrow\,A^2\,.
\end{equation}

\bibliographystyle{JHEP}
\bibliography{20230210}

\end{document}